\documentclass[sigconf]{acmart}

\AtBeginDocument{%
  \providecommand\BibTeX{{%
    \normalfont B\kern-0.5em{\scshape i\kern-0.25em b}\kern-0.8em\TeX}}}

\usepackage{booktabs}
\usepackage{array}
\usepackage{balance} 
\usepackage{multirow}
\usepackage[normalem]{ulem}
\usepackage{color}
\definecolor{lightgray}{RGB}{215,215,215}
\usepackage{colortbl}
\usepackage{xcolor}
\useunder{\uline}{\ul}{}
\usepackage{subfigure}
\usepackage{algorithm}  
\usepackage{algorithmicx}  
\usepackage[noend]{algpseudocode}  
\usepackage{stackengine}

\usepackage{amsmath}  
\usepackage{enumitem}
\usepackage{tabularx}
\usepackage[utf8]{inputenc}
\usepackage[english]{babel}
\usepackage{amsthm}
\usepackage{bm}
\usepackage{threeparttable}
\newcommand{\ie}{\emph{i.e., }}
\newcommand{\eg}{\emph{e.g., }}

\newcommand\xrowht[2][0]{\addstackgap[.5\dimexpr#2\relax]{\vphantom{#1}}}

\newlength\myindent
\floatname{algorithm}{Algorithm}

\copyrightyear{2024}
\acmYear{2024}
\setcopyright{acmlicensed}\acmConference[SIGIR '24]{Proceedings of the 47th
International ACM SIGIR Conference on Research and Development in
Information Retrieval}{July 14--18, 2024}{Washington, DC, USA}
\acmBooktitle{Proceedings of the 47th International ACM SIGIR Conference on
Research and Development in Information Retrieval (SIGIR '24), July 14--18,
2024, Washington, DC, USA}
\acmDOI{10.1145/3626772.3657719}
\acmISBN{979-8-4007-0431-4/24/07}

\begin{document}


\title{Diffusion Models for Generative Outfit Recommendation}

\author{Yiyan Xu}
\email{yiyanxu24@gmail.com}
\affiliation{
\institution{University of Science and Technology of China}
\country{China}
\city{Hefei}
}

\author{Wenjie Wang}
\authornote{Corresponding authors. This work is supported by the National Key Research and Development Program of China (2022YFB3104701), the National Natural Science Foundation of China (62272437), and the CCCD Key Lab of Ministry of Culture and Tourism.}
\email{wenjiewang96@gmail.com}
\affiliation{
\institution{National University of Singapore}
\country{Singapore}
\city{Singapore}
}

\author{Fuli Feng}
\authornotemark[1]
\email{fulifeng93@gmail.com}
\affiliation{
\institution{University of Science and Technology of China}
\country{China}
\city{Hefei}
}

\author{Yunshan Ma}
\email{yunshan.ma@u.nus.edu}
\affiliation{
\institution{National University of Singapore}
\country{Singapore}
\city{Singapore}
}

\author{Jizhi Zhang}
\email{cdzhangjizhi@mail.ustc.edu.cn}
\affiliation{
\institution{University of Science and Technology of China}
\country{China}
\city{Hefei}
}

\author{Xiangnan He}
\email{xiangnanhe@gmail.com}
\affiliation{
\institution{University of Science and Technology of China}
\country{China}
\city{Hefei}
}

\renewcommand{\shortauthors}{Yiyan Xu et al.}

\begin{abstract}

Outfit Recommendation (OR) in the fashion domain has evolved through two stages: Pre-defined Outfit Recommendation and Personalized Outfit Composition. 
However, both stages are constrained by existing fashion products, limiting their effectiveness in addressing users' diverse fashion needs. Recently, the advent of AI-generated content provides the opportunity for OR to transcend these limitations, showcasing the potential for personalized outfit generation and recommendation. 

To this end, we introduce a novel task called Generative Outfit Recommendation (GOR), aiming to generate a set of fashion images and compose them into a visually compatible outfit tailored to specific users. The key objectives of GOR lie in the high fidelity, compatibility, and personalization of generated outfits. To achieve these, we propose a generative outfit recommender model named DiFashion, which empowers exceptional diffusion models to accomplish the parallel generation of multiple fashion images. To ensure three objectives, we design three kinds of conditions to guide the parallel generation process and adopt Classifier-Free-Guidance to enhance the alignment between the generated images and conditions. 
We apply DiFashion on both personalized Fill-In-The-Blank and GOR tasks and conduct extensive experiments on iFashion and Polyvore-U datasets. The quantitative and human-involved qualitative evaluation demonstrate the superiority of DiFashion over competitive baselines.

\end{abstract}


\maketitle

\section{Introduction}
\label{sec:introduction}

Fashion significantly influences daily human life, reflecting individual personality and aesthetics. However, many consumers often struggle to assemble diverse fashion elements into a compatible outfit, largely due to limited fashion expertise~\cite{bettaney2020og,celikik2021outfit,chen2021tops}. As a result, Outfit Recommendation (OR) has emerged as an essential task in the fashion domain, offering personalized outfit-level fashion advice. By suggesting well-matched outfits, OR streamlines the fashion decision-making process, saving consumers from the extra time and effort needed to coordinate various fashion products.

\begin{figure}[t]
\setlength{\abovecaptionskip}{0.1cm}
\setlength{\belowcaptionskip}{-0.3cm}
\centering
\includegraphics[scale=0.5]{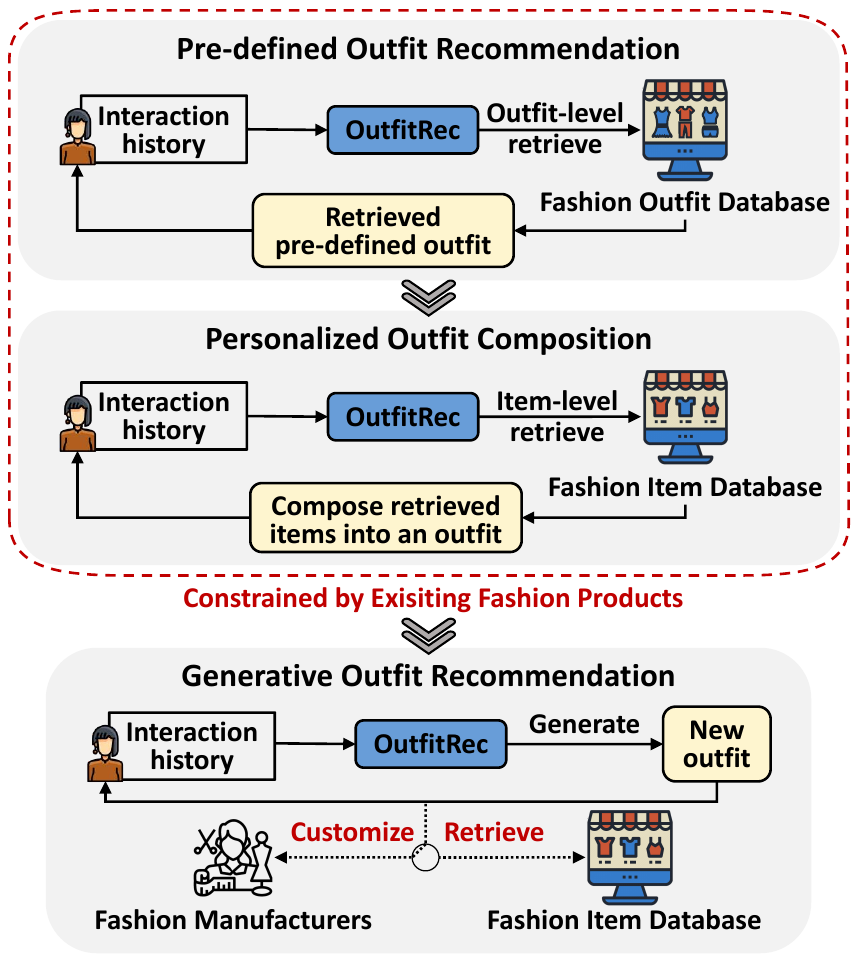}
\caption{The evolution of Outfit Recommendation. 
Beyond existing fashion products, GOR aims to generate a set of fashion products as a compatible and personalized outfit.
}
\label{fig:intro_outfitrec}
\end{figure}

The evolution of OR, as illustrated at the top of Figure~\ref{fig:intro_outfitrec}, unfolds through two stages:
\begin{itemize}[leftmargin=*]
    \item \textbf{Pre-defined Outfit Recommendation (POR)}: OR is initially framed as an outfit-level retrieval task, aiming to retrieve existing well-matched outfits for users~\cite{li2020hfgn,lu2019hfn,lu2021personalized,zhan2021A3FKG,dong2019capsule}. However, its effectiveness is hindered by the limited quantity and diversity of available pre-defined outfits, thus making it insufficient for meeting users' diverse fashion preferences~\cite{ding2023og}.

    \item \textbf{Personalized Outfit Composition}: This task focuses on retrieving a bundle of fashion products at the item level and composing them into visually harmonious outfits~\cite{han2017bilstm,chen2019pog,ding2023og}. By composing new, personalized, and compatible outfits, this task partly alleviates the diversity constraints of pre-defined outfit recommendation. However, existing fashion products might not fully align with users' ideal expectations, especially in details such as colors, patterns, cuts, and designs~\cite{yu2019personalized,jain2019text}.
\end{itemize}

With the boom of AI-generated content (AIGC)~\cite{dhariwal2021diffusion,rombach2022high,brown2020language}, OR shows the potential to evolve beyond existing fashion products, enabling customized generation and composition of fashion products.
In light of this, we propose a new task called Generative Outfit Recommendation (GOR). As shown in Figure~\ref{fig:intro_outfitrec}, GOR aims to generate a set of new personalized fashion products to compose a visually compatible outfit catering to specific users' fashion tastes. With the generated new outfits, GOR could readily retrieve similar products in the fashion item database~\cite{yu2019personalized,wu2021fashion,yuan2021conversational}
or turn to fashion manufacturers\footnote{https://home.knitup.io.} for customization~\cite{sharma2021development}.
This generative task holds the promise to provide truly tailored and diverse fashion outfits that closely align with users' ideal fashion requirements. 

To effectively implement GOR using generative models, three key criteria are essential: 1) \textit{High fidelity}, which requires the generated images to accurately depict the details of fashion products, 2)  \textit{Compatibility}, ensuring the generated fashion images harmonize cohesively within an outfit, and 3) \textit{Personalization}, which demands alignment of the generated outfits with specific user preferences. Diffusion Models (DMs), known for their state-of-the-art capabilities in image synthesis, hold the potential to generate high-quality fashion images~\cite{ho2020denoising,chen2023pixartalpha,rombach2022high}. Previous studies have employed DMs to generate a single fashion image from text prompts and conditions of other modalities~\cite{sun2023sgdiff,kong2023leveraging,cao2023image,yan2023fashiondiff,guo2023ai}. 
However, GOR requires generating multiple fashion images with internal compatibility for outfit composition and recommendation, emphasizing the alignment of generated outfits with individual fashion tastes captured from user information (\eg interaction history and user features).

To generate high-fidelity, compatible, and personalized outfits for users, we propose \textit{DiFashion}, a generative outfit recommender model adapted from DMs. Generally, DiFashion comprises two critical processes: gradually corrupting fashion images within the same outfit with Gaussian noise in the forward process, followed by a conditional denoising process to generate these images in parallel during the reverse phase.
Three conditions, \ie category prompt, mutual condition, and history condition, are the keys to guiding the parallel generation process to pursue the three criteria of GOR.
Specifically, 1) for \textit{high fidelity}, DiFashion employs category prompts to ensure category consistency and adopts Classifier-Free-Guidance~\cite{ho2022classifier,brooks2023instructpix2pix} for the three conditions to enhance image quality and alignment between the generated images and conditions,
2) to ensure \textit{compatibility}, a mutual encoder is designed to encode fashion images within the same outfit at different noise levels into compatibility information, serving as the mutual condition, 
and 3) for \textit{personalization}, DiFashion includes a history encoder that leverages users' historical interactions with fashion products, capturing their personalized tastes as the history condition.

\begin{figure}[t]
\setlength{\abovecaptionskip}{0.03cm}
\setlength{\belowcaptionskip}{-0.7cm}
\centering
\includegraphics[scale=0.43]{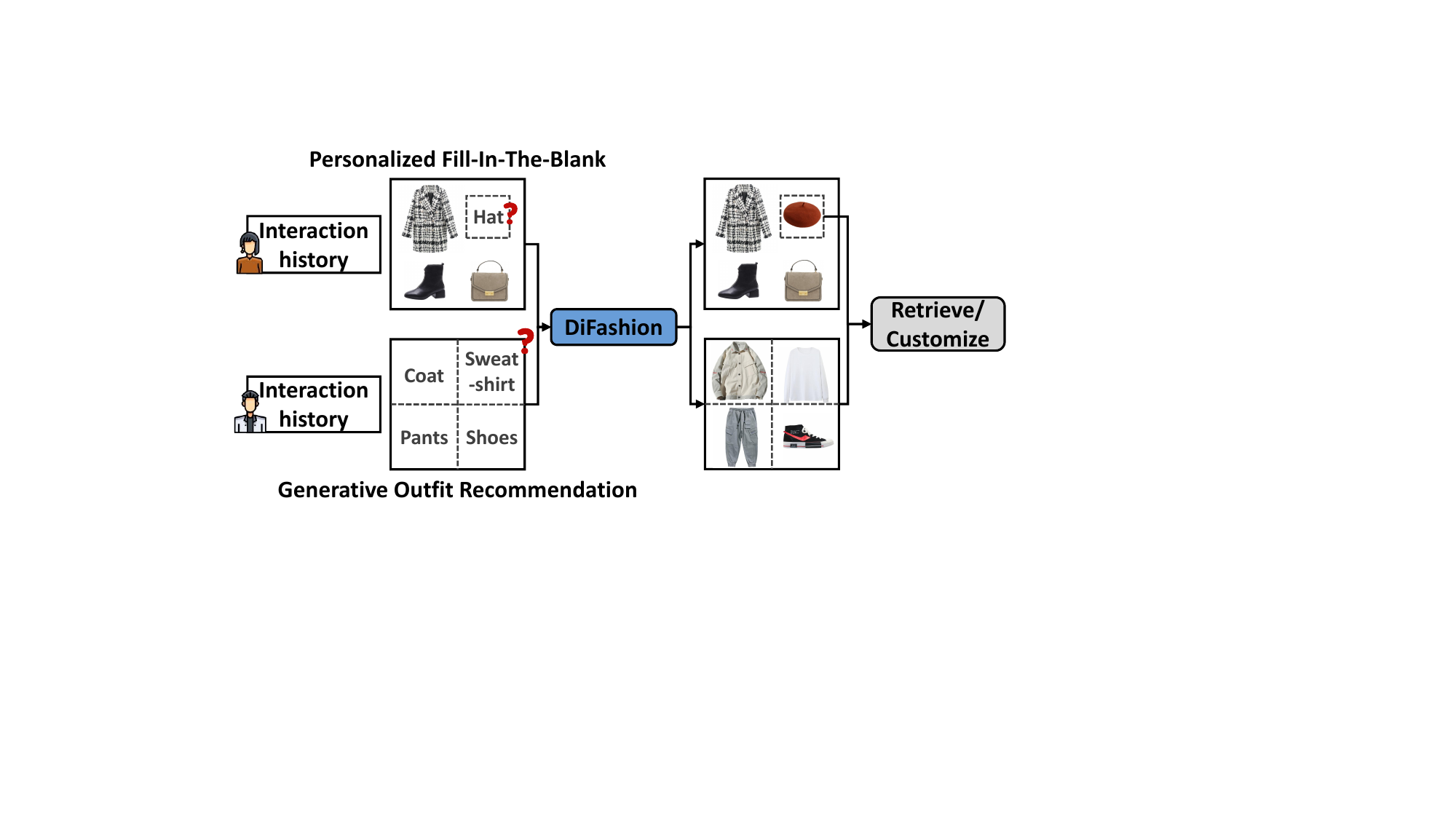}
\caption{Demonstration of DiFashion for personalized Fill-In-The-Blank and generative outfit recommendation tasks.}
\label{fig:demo}
\end{figure}

We instantiate DiFashion on two tasks: Personalized Fill-In-The-Blank (PFITB) and GOR tasks (see illustration in Figure~\ref{fig:demo}). 
Given users' interaction history and designated categories, the PFITB task involves generating a fashion product that seamlessly complements an incomplete outfit, while the GOR task requires producing a whole compatible outfit.
We conduct extensive experiments on iFashion~\cite{chen2019pog} and Polyvore-U~\cite{lu2019hfn} datasets and compare DiFashion with various baselines, including generative models and retrieval-based models, demonstrating the superiority of our proposed DiFashion in both PFITB and GOR tasks. We have released our code and data at \url{https://github.com/YiyanXu/DiFashion}.

In summary, the key contributions of this work are as follows:
\vspace{-0.1cm}
\begin{itemize}[leftmargin=*]
    \item We propose a new generative outfit recommendation task, using Generative AI to create personalized outfits for users. This initiative pioneers a promising avenue for outfit recommendation and contributes to a more personalized fashion landscape. 
    \item We present DiFashion, a generative outfit recommender model, which adepts at the parallel generation of multiple fashion images, skillfully pursuing various generation objectives. 
    \item We conduct substantial experiments on iFashion and Polyvore-U datasets, where both quantitative and human-involved qualitative evaluation demonstrate the remarkable performance of DiFashion on PFITB and GOR tasks. 
\end{itemize}


\section{Preliminary}
\label{sec:preliminaries}

\begin{figure*}
\setlength{\abovecaptionskip}{0.0cm}
\setlength{\belowcaptionskip}{-0.42cm}
\centering
\includegraphics[scale=0.632]{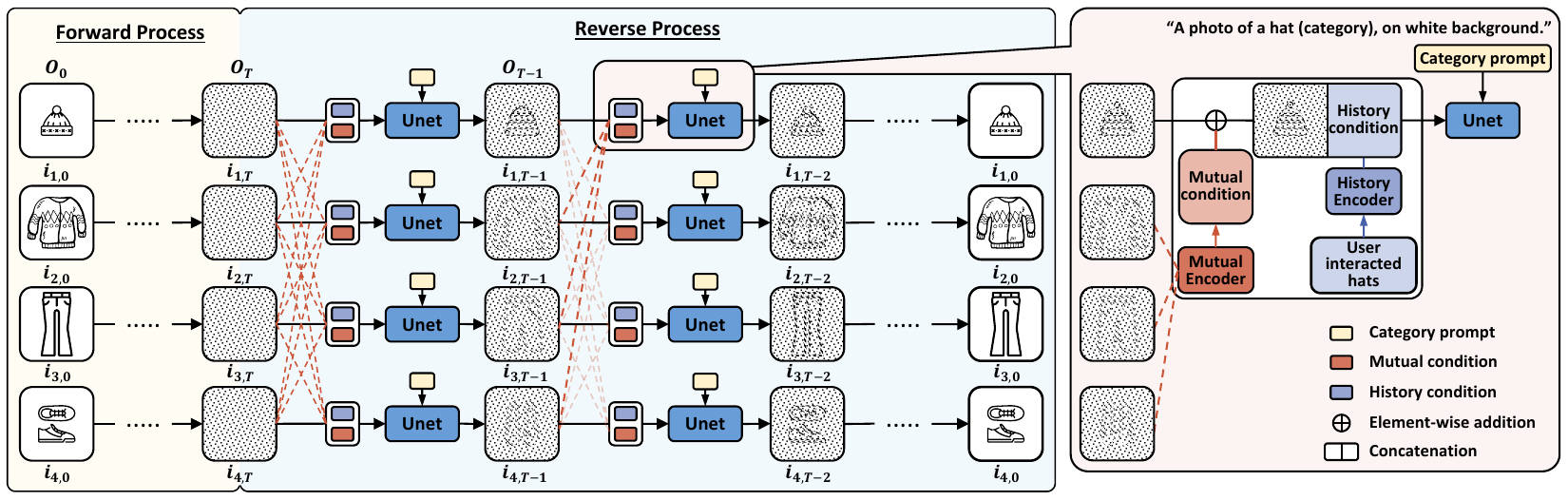}
\caption{An overview of DiFashion: it gradually corrupts outfit images with Gaussian noise in the forward process, followed by a parallel conditional denoising process to reconstruct these images. The denoising process is guided by three conditions: category prompt, mutual condition, and history condition.}
\label{fig:method}
\end{figure*}

DMs employ Markov chains with $T$ diffusion steps for image synthesis, involving the forward and reverse processes.

\vspace{3pt}
\noindent\textbf{$\bullet $ Forward process}. Given an image $\bm{x_0}\sim q(\bm{x_0})$, the forward diffusion process corrupts images by adding Gaussian noises:
\begin{equation}
\setlength{\abovedisplayskip}{2pt}
\setlength{\belowdisplayskip}{2pt}
    q(\bm{x_t}|\bm{x_{t-1}})=\mathcal{N}(\bm{x_t};\sqrt{1-\beta_t}\bm{x_{t-1}},\beta_t\bm{I}),
\end{equation}
where $t\in\{1,\dots,T\}$ denotes the diffusion step, and $\beta_t\in(0,1)$ controls the noise scale added at each step $t$. 
As $T\rightarrow\infty$, the image $\bm{x_T}$ converges to standard Gaussian noise $\bm{x_T}\sim\mathcal{N}(0,\bm{I})$~\cite{ho2020denoising}. 

\vspace{3pt}
\noindent\textbf{$\bullet $ Reverse process}. Starting from a nearly pure noisy image $\bm{x}_T$, DMs iteratively eliminate the noises by the denoising transition:
\begin{equation}
\setlength{\abovedisplayskip}{2pt}
\setlength{\belowdisplayskip}{2pt}
    p_\theta(\bm{x}_{t-1}|\bm{x}_t)=\mathcal{N}(\bm{x}_{t-1};\bm{\mu}_\theta(\bm{x}_t,t),\bm{\Sigma}_\theta(\bm{x}_t,t)),
\label{eq:diff_reverse}
\end{equation}
where $\bm{\mu}_\theta$ and $\bm{\Sigma}_\theta$ represent the approximated mean and covariance of the reversed Gaussian distribution parameterized by a neural network (\eg U-Net~\cite{ho2020denoising} or Transformer~\cite{peebles2023scalable}). To maintain training stability, the learning of $\bm{\Sigma}_\theta$ is commonly ignored~\cite{ho2020denoising}, while the denoising transition mean $\bm{\mu}_\theta$ can be further factorized as:
\begin{equation}
\setlength{\abovedisplayskip}{0pt}
\setlength{\belowdisplayskip}{2pt}
    \bm{\mu}_\theta(\bm{x_t},t)=\dfrac{1}{\sqrt{\alpha_t}}\left(\bm{x_t}-\dfrac{1-\alpha_t}{\sqrt{1-\bar{\alpha}_t}}\bm{\epsilon}_\theta(\bm{x_t},t)\right),
\label{eq:diff_mean}
\end{equation}
where $\alpha_t=1-\beta_t$, $\bar{\alpha}_t=\prod_{t'=1}^t\alpha_{t'}$ and $\bm{\epsilon}_\theta$ learns to predict the source noise $\bm{\epsilon}\sim\mathcal{N}(0,\bm{I})$ determining $\bm{x_t}$ from $\bm{x_0}$~\cite{luo2022understanding}.

\vspace{3pt}
\noindent\textbf{$\bullet $ Optimization}. The denoising neural network $\theta$ can be trained using the following simplified objective~\cite{ho2020denoising}:
\begin{equation}
\setlength{\abovedisplayskip}{2pt}
\setlength{\belowdisplayskip}{2pt}
    \mathcal{L}_\theta=\mathbb{E}_{t,\bm{\epsilon}\sim\mathcal{N}(0,\bm{I})}\left[\parallel\bm{\epsilon}-\bm{\epsilon}_\theta(\bm{x_t},t)\parallel_2^2\right],
\end{equation}
with $t$ uniformly sampled from $\{1,\dots,T\}$.

\vspace{3pt}
\noindent\textbf{$\bullet $ Generation}. After training $\theta$, DMs randomly sample a pure noise $\bm{x_T}\sim\mathcal{N}(0,\bm{I})$ as the initial state, and conduct the generation process $\bm{x_T}\rightarrow\bm{x_{T-1}}\rightarrow\cdots\rightarrow\bm{x_0}$ via Eq.~(\ref{eq:diff_reverse}) and Eq.~(\ref{eq:diff_mean}) iteratively.

\vspace{-0.12cm}

\section{Generative Outfit Recommendation}
\label{sec:method}

\subsection{Task Formulation}
\label{sec:task_formulation}
The GOR task aims to synthesize a set of novel fashion product images to compose a visually compatible outfit that aligns with personalized user preferences. 
Formally, given user information $\bm{u}$ (\eg interaction history and user features), GOR generates an outfit $\bm{O}=\{\bm{i}_k\}_{k=1}^n$, where $\bm{i}_k$ denotes an individual fashion item, to satisfy three criteria: high fidelity, compatibility, and personalization. We can define GOR as the following optimization problem:
\begin{subequations}
\setlength{\abovedisplayskip}{-8pt}
\setlength{\belowdisplayskip}{1pt}
\begin{gather}
    \bm{u},\emptyset\stackrel{\theta}{\longrightarrow} \bm{O}=\{\bm{i}_k\}_{k=1}^n, \label{eq:og_def_process} \\
    \text{s.t.}~\bm{i}_k=\arg\max_{\bm{i}}~P_\theta(\bm{i} | \bm{O}_k', \bm{u}),~k=1,\dots,n.
\label{eq:og_def_constraint}
\end{gather}
\label{eq:og_def}
\end{subequations} 
Here, user information $\bm{u}$, along with the empty set $\emptyset$, serves as inputs for the generative model with parameters $\theta$, leading to the generation of the outfit $\bm{O}$. The generation process is constrained by Eq.~(\ref{eq:og_def_constraint}), where $P_\theta(\cdot)$ signifies the conditional probability of generating each item. The model parameters $\theta$ are optimized to generate the most compatible match item $\bm{i}_k$ for each incomplete outfit $\bm{O}_k'=\bm{O}\setminus\{\bm{i}_k\}$ while aligning with user preferences.

However, this optimization problem is intractable due to the lack of a well-defined starting point and the constraints continually evolve during the optimization process. 
An intuitive solution~\cite{han2017bilstm,chen2019pog} simplifies the constraints into $\bm{i}_k=\arg\max_{\bm{i}}P_\theta(\bm{i}|\bm{i}_1,\cdots,\bm{i}_{k-1})$ by modeling outfits as ordered sequences, solely considering influences from pre-ordered items to guide the generation of $\bm{i}_k$, thus making the optimization tractable. Nevertheless, it is not the ideal solution, as previously generated items in the sequence are not influenced by subsequent ones~\cite{cui2019dressing,ding2023og}, deviating from the original optimization objective in Eq.~(\ref{eq:og_def}).

DMs generate images in an iterative manner, making it possible to solve the optimization problem via a multi-step generation process. Firstly, let us consider the parallel generation of $n$ images without delving into implementation details. Starting from multiple pure noises $\bm{O}_T=\{\bm{i}_{k,T}\}_{k=1}^{n}$ where $\bm{i}_{k,T}\sim\mathcal{N}(0,\bm{I})$, DMs
gradually repeat the generation process $\bm{O}_T\rightarrow \bm{O}_{T-1}\rightarrow\cdots\rightarrow \bm{O}_0$, resulting in $T$ outfits
at different noise levels. Even though the accurate incomplete outfit $\bm{O}_k'$ in Eq.~(\ref{eq:og_def_constraint}) is unreachable, $\bm{O}_{k,t}'=\bm{O}_t\setminus\{\bm{i}_{k,t}\}$ could serve as a surrogate one for the generation of $\bm{i}_{k,t-1}$ at each diffusion step $t\in\{1,\dots,T\}$.  Formally, the optimization problem in Eq.~(\ref{eq:og_def}) could be approximately expressed as:

\begin{subequations}
\setlength{\abovedisplayskip}{-7pt}
\setlength{\belowdisplayskip}{-7pt}
\begin{gather}
    \bm{u},\bm{O}_T=\{\bm{i}_{k,T}\}_{k=1}^{n}\stackrel{\theta}{\longrightarrow} \bm{O}_0=\{\bm{i}_{k,0}\}_{k=1}^{n},\label{eq:og_diff_def_process} \\
    \text{s.t.}~\bm{i}_{k,t-1}=\arg\max_{\bm{i}}~P_\theta(\bm{i} | \bm{i}_{k,t}, \bm{O}_{k,t}', \bm{u}),\label{eq:og_diff_def_constraint}\\
    \text{where}~k=1,\dots,n~\text{and}~t=1,\dots,T.\nonumber
\end{gather}
\label{eq:og_diff_def}
\end{subequations}

\noindent Note that the initial state $\bm{O}_T$ is equivalent to $\emptyset$ in Eq.~(\ref{eq:og_def_process}), as each element $\bm{i}_{k,T}$ is sampled from a standard Gaussian noise and contains no information.

\vspace{-0.2cm}
\subsection{DiFashion}
To generate a set of compatible fashion images based on user information, as depicted in Figure~\ref{fig:method}, DiFashion gradually corrupts multiple fashion images to nearly pure noises in the forward process and subsequently conducts the parallel reverse reconstruction of these images by the U-Net. During the reverse process, DiFashion considers three crucial input conditions: 1) \textit{category prompt} ensures the generated fashion images align with designated categories, 2) \textit{mutual condition} guarantees internal compatibility
within generated outfits and 3) \textit{history condition} ensures the generated outfits closely match user preferences. 
These conditions collaborate to guide the reverse process, resulting in visually appealing outfits tailored to users' unique fashion tastes. Note that only four items within the outfit are illustrated in Figure~\ref{fig:method} for brevity, while it is simple for DiFashion to handle more items simultaneously.

\vspace{-0.16cm}
\subsubsection{\textbf{Diffusion Processes}}
\label{sec:forward_reverse}
DiFashion involves two crucial diffusion processes: the forward process corrupts outfit images by adding Gaussian noises, and the reverse process learns to reconstruct these images in parallel, both executed in a low-dimensional latent space instead of the original pixel space. Following Stable Diffusion (SD)~\cite{rombach2022high}, we utilize a pre-trained autoencoder with encoder $\mathcal{E}$ and decoder $\mathcal{D}$ to compress the outfit $\bm{O}_0=\{\bm{i}_{k,0}\}_{k=1}^n$ into the latent space $\mathcal{E}(\bm{O}_0)=\{\mathcal{E}(\bm{i}_{k,0})\}_{k=1}^n$. After the reverse process, the outfit is reverted to the pixel space using the decoder $\mathcal{D}$. For brevity, we use the abbreviation $\bm{O}_0$ and $\bm{i}_{k,0}$ for the diffusion processes, omitting details about $\mathcal{E}$ and $\mathcal{D}$, as shown in Figure~\ref{fig:method}.

\vspace{3pt}
\noindent\textbf{$\bullet $ Forward Process.} Given an outfit $\bm{O}_0=\{\bm{i}_{k,0}\}_{k=1}^n$, where $\bm{i}_{k,0}$ denotes a fashion product within the outfit, the forward transition of the outfit is performed independently on each image, defined as:
\begin{equation}
\setlength{\abovedisplayskip}{-1pt}
\setlength{\belowdisplayskip}{1pt}
\begin{aligned}
    q(\bm{O}_t|\bm{O}_{t-1}) & =\prod_{k=1}^nq(\bm{i}_{k,t}|\bm{i}_{k,t-1}) \\
    & =\prod_{k=1}^n\mathcal{N}(\bm{i}_{k,t};\sqrt{1-\beta_t}\bm{i}_{k,t-1},\beta_t\bm{I}),
\end{aligned}
\end{equation}
where $\beta_t\in(0,1)$ controls the noise scales added at each diffusion step $t\in\{1,\dots,T\}$. As $T\rightarrow\infty$, the outfit $\bm{O}_T$ converges to a set of standard Gaussian noises~\cite{ho2020denoising}.

\vspace{3pt}
\noindent\textbf{$\bullet $ Reverse Process.} Starting from the nearly pure noises $\bm{O}_T$, the reverse process gradually reconstructs fashion images within the outfit in parallel, through the following denoising transition step:
\begin{equation}
\setlength{\abovedisplayskip}{-1pt}
\setlength{\belowdisplayskip}{1pt}
\begin{aligned}
    p_\theta(\bm{O}_{t-1}|\bm{O}_t) &=\prod_{k=1}^np_\theta(\bm{i}_{k,t-1}|\bm{O}_t),\\
    &=\prod_{k=1}^n\mathcal{N}(\bm{i}_{k,t-1};\bm{\mu}_\theta(\bm{O}_t,t),\bm{\Sigma}_\theta(\bm{O}_t,t)),
\end{aligned}
\end{equation}
where $\bm{\mu}_\theta$ and $\bm{\Sigma}_\theta$ denote the Gaussian parameters outputted by the U-Net with learnable parameters $\theta$. The outfit at step $t$, $\bm{O}_t=\{\bm{i}_{k,t}\}\cup \bm{O}_{k,t}'$, comprises the current noisy image $\bm{i}_{k,t}$ and the incomplete noisy outfit $\bm{O}_{k,t}'$, where the latter contains compatibility information to guide the denoising process.
Additionally, for category consistency and personalization, when handling $\bm{i}_{k,t}$ with category $c_k$, we introduce category prompt $\bm{t}_{c_k}$ and user's interaction history $\bm{u}$ into the reverse transition step:
\begin{equation}
\setlength{\abovedisplayskip}{0pt}
\setlength{\belowdisplayskip}{3pt}
    p_\theta(\bm{O}_{t-1}|\bm{O}_t,\bm{t}_{c_k},\bm{u})=\prod_{k=1}^np_\theta(\bm{i}_{k,t-1}|\bm{i}_{k,t},\bm{t}_{c_k},\bm{O}_{k,t}',\bm{u}).
\label{eq:outfit_reverse}
\end{equation}
Here, $p_\theta(\cdot)$ is analogous to $P_\theta(\cdot)$ in Eq.~(\ref{eq:og_diff_def}), representing the conditional probability of generating $\bm{i}_{k,t-1}$ as the most harmonious match item given the specified conditions.

\vspace{-0.13cm}
\subsubsection{\textbf{Condition Encoders}}
As illustrated in Figure~\ref{fig:method}, three conditions are employed in the reverse process. For the category prompt $\bm{t}_{c_k}$, such as \textit{``A photo of a hat, on white background"}, we follow the text conditioning mechanism of SD~\cite{rombach2022high}, which is omitted in the figure for brevity. 
Additionally, to capture compatibility information and user preferences,
we introduce dedicated encoders for each incomplete outfit $\bm{O}_{k,t}'$ and user's interaction history $\bm{u}$ to obtain mutual and history conditions:

\vspace{3pt}
\noindent\textbf{$\bullet $ Mutual Encoder.} 
During the reverse process, the outfit $\bm{O}_t$ could construct the incomplete outfit $\bm{O}_{k,t}'$
to provide compatibility information for the generation of $\bm{i}_{k,t-1}$ at each diffusion step $t$. For instance, as shown in Figure~\ref{fig:method}, in the generation of the hat $\bm{i}_{1,T-2}$, other fashion items $\bm{O}_{1,T-1}'=\{$sweater $\bm{i}_{2,T-1}$, jeans $\bm{i}_{3,T-1}$, shoes $\bm{i}_{4,T-1}\}$ offer compatibility information. To better capture this information, we compute the averaged influence for each incomplete outfit $\bm{O}_{k,t}'$ in Eq.~(\ref{eq:outfit_reverse}) and feed it into a Multi-Layer Perceptron (MLP) $f_\phi(\cdot)$ to obtain the mutual condition $\bm{m}_{k,t}$:
\begin{equation}
\setlength{\abovedisplayskip}{-2pt}
\setlength{\belowdisplayskip}{3pt}
\bm{m}_{k,t}=f_\phi\left(Avg(\bm{O}_{k,t}')\right)=f_\phi\left(\frac{1}{n-1}\sum_{v\neq k}\bm{i}_{v,t}\right),
\label{eq:mutual_cond}
\end{equation}
where $\phi$ is the learnable parameters of the MLP. This condition guides the denoising process through element-wise addition:
\begin{equation}
\setlength{\abovedisplayskip}{2pt}
\setlength{\belowdisplayskip}{2pt}
    \bm{i}_{k,t}^{\text{mutual}}=(1-\eta)\cdot\bm{i}_{k,t}+\eta\cdot\bm{m}_{k,t},
\label{eq:mutual}
\end{equation}
where $\eta$ denotes the mutual influence ratio. In this way, the mutual condition enhances the compatibility between $\bm{i}_{k,t-1}$ and other fashion items. Importantly, all the mutual conditions $\{\bm{m}_{k,t}\}_{k=1}^n$ collaborate to ensure internal compatibility within the outfit $\bm{O}_{t-1}$.

\vspace{3pt}
\noindent\textbf{$\bullet $ History Encoder.} User preferences are captured from their interaction history within each category. For instance, during the generation of the hat (see Figure~\ref{fig:method}), user-interacted hats provide personalized information at each diffusion step. Formally, to generate $\bm{i}_{k,t-1}$ with category $c_k$ in Eq.~(\ref{eq:outfit_reverse}), user's interaction history $\bm{u}$ is categorized into $\bm{u}_{c_k}=\{\bm{i}_{c_k}^r\}_{r=1}^m$, indicating $m$ interactions with fashion products in category $c_k$. Utilizing the same pre-trained encoder $\mathcal{E}$ mentioned in Section \ref{sec:forward_reverse}, we compress these history images into latent space, denoted as $\mathcal{E}(\bm{u}_{c_k})=\{\mathcal{E}(\bm{i}_{c_k}^r)\}_{r=1}^m$. Next, we average these latent representations as the history condition:
\begin{equation}
\setlength{\abovedisplayskip}{1pt}
\setlength{\belowdisplayskip}{2pt}
    \bm{h}_{c_k} = Avg(\mathcal{E}(\bm{u}_{c_k}))=\frac{1}{m}\sum_{r=1}^m\mathcal{E}(\bm{i}_{c_k}^r).
\label{eq:hist_cond}
\end{equation}
Following \cite{brooks2023instructpix2pix}, we expand the first convolutional layer of the U-Net by introducing additional input channels, functioning as a built-in history encoder initialized with zero. This modification allows DiFashion to integrate the history condition by concatenating it with $\bm{i}_{k,t}^{\text{mutual}}$ in Eq.~(\ref{eq:mutual}), thereby forming the U-Net input $[\bm{i}_{k,t}^{\text{mutual}},\bm{h}_{c_k}]$.

In summary, DiFashion incorporates three conditions via different mechanisms, guiding the denoising process in the U-Net.

\subsubsection{\textbf{Training}}
In essence, DiFashion is built on SD to integrate three conditions for multiple-image synthesis. By incorporating category prompt $\bm{t}_{c_k}$, mutual condition $\bm{m}_{k,t}$ and history condition $\bm{h}_{k}$, the U-Net is optimized to predict the noises $\bm{\epsilon}_k$ added in the forward process by minimizing the following objective~\cite{ho2020denoising}:
\begin{equation}
\small
\setlength{\abovedisplayskip}{1pt}
\setlength{\belowdisplayskip}{2pt}
    \mathcal{L}_{\theta,\phi}=\frac{1}{n}\sum_{k=1}^n\mathbb{E}_{t,\bm{\epsilon}_k\sim\mathcal{N}(0,\bm{I})}\left[\parallel\bm{\epsilon}_k-\bm{\epsilon}_{\theta,\phi}(\bm{i}_{k,t},\bm{t}_{c_k},\bm{m}_{k,t},\bm{h}_{c_k},t)\parallel_2^2\right].
\end{equation}

In conditional image synthesis, Classifier-Free-Guidance (CFG)
~\cite{ho2022classifier} is commonly employed to enhance image fidelity and ensure the generated images align closely with input conditions. Specifically, in classical text-to-image tasks, for a noisy image $\bm{x}_t$, DMs are trained under both conditional scene $txt$ and unconditional scene $\emptyset$, where $\emptyset$ denotes a fixed tensor without conditional information. The predicted noises under these scenes are denoted as $\bm{\epsilon}_\theta(\bm{x}_t,txt,t)$ and $\bm{\epsilon}_\theta(\bm{x}_t,\emptyset,t)$, respectively. During inference, the two estimates are combined to guide the generation process:
\begin{equation}
\setlength{\abovedisplayskip}{3pt}
\setlength{\belowdisplayskip}{3pt}
\bm{\tilde{\epsilon}}_\theta(\bm{x}_t,txt,t)=\bm{\epsilon}_\theta(\bm{x}_t,\emptyset,t)+s\cdot[\bm{\epsilon}_\theta(\bm{x}_t,txt,t)-\bm{\epsilon}_\theta(\bm{x}_t,\emptyset,t)],
\end{equation}
where $s\ge 1$ represents the guidance scale of the condition.

In the DiFashion scenario, three conditions guide the generation process: category prompt $\bm{t}_{c_k}$, mutual condition $\bm{m}_{k,t}$ and history condition $\bm{h}_{c_k}$. 
Following \cite{brooks2023instructpix2pix}, we extend CFG for three conditions, resulting in the modified prediction for the inference phase:
\begin{equation}
\setlength{\abovedisplayskip}{3pt}
\setlength{\belowdisplayskip}{3pt}
\begin{aligned}    &\bm{\tilde{\epsilon}}_{\theta,\phi}(\bm{i}_{k,t},\bm{t}_{c_k},\bm{m}_{k,t},\bm{h}_{c_k},t) \\
=~&\bm{\epsilon}_{\theta,\phi}(\bm{i}_{k,t},\emptyset,\emptyset,\emptyset,t) \\
+~&s_t\cdot[\bm{\epsilon}_{\theta,\phi}(\bm{i}_{k,t},\bm{t}_{c_k},\emptyset,\emptyset,t) - \bm{\epsilon}_{\theta,\phi}(\bm{i}_{k,t},\emptyset,\emptyset,\emptyset,t)] \\
+~&s_m\cdot[\bm{\epsilon}_{\theta,\phi}(\bm{i}_{k,t},\bm{t}_{c_k},\bm{m}_{k,t},\emptyset,t)-\bm{\epsilon}_{\theta,\phi}(\bm{i}_{k,t},\bm{t}_{c_k},\emptyset,\emptyset,t)] \\
+~&s_h\cdot[\bm{\epsilon}_{\theta,\phi}(\bm{i}_{k,t},\bm{t}_{c_k},\bm{m}_{k,t},\bm{h}_{c_k},t)-\bm{\epsilon}_{\theta,\phi}(\bm{i}_{k,t},\bm{t}_{c_k},\bm{m}_{k,t},\emptyset,t)],
\end{aligned}
\label{eq:cfg_difashion}
\end{equation}
where $s_t$, $s_m$ and $s_h$ denote the guidance scales of category prompt, mutual condition, and history condition, respectively. To support the CFG approach for inference, DiFashion randomly masks these three conditions with specific ratios during training, ensuring the model parameters are trained under all scenes in Eq.~(\ref{eq:cfg_difashion}).

\subsubsection{\textbf{Inference}}
With well-trained model parameters, DiFashion is capable of performing the following two tasks:

\vspace{3pt}
\noindent\textbf{$\bullet$ PFITB.} 
This task involves synthesizing a personalized new fashion product that seamlessly complements an incomplete outfit. Formally, given an incomplete outfit $\bm{O}_n'=\{\bm{i}_k\}_{k=1}^{n-1}$, a designated category $c_n$, category prompt $\bm{t}_{c_n}$ and user's interaction history $\bm{u}_{c_n}$, DiFashion aims to generate the missing product $\bm{i}_n$. Note that due to the existence of the exact incomplete outfit $\bm{O}_n'$, it is unnecessary to use the surrogate $\bm{O}_{n,t}'$ to obtain the mutual condition for each generation step $t$.  Starting from a pure noise $\bm{i}_{n,T}\sim\mathcal{N}(0,\bm{I})$, DiFashion computes the mutual condition $\bm{m}_n$ via Eq.~(\ref{eq:mutual_cond}) and the history condition $\bm{h}_{c_n}$ via Eq.~(\ref{eq:hist_cond}). With these conditions, it predicts the noise $\bm{\tilde{\epsilon}}_{\theta,\phi}(\bm{i}_{n,t},\bm{t}_{c_n},\bm{m}_{n},\bm{h}_{c_n},t)$ at each diffusion step via Eq.~(\ref{eq:cfg_difashion}), and iteratively removes the noise $\bm{i}_{n,t}\rightarrow\bm{i}_{n,t-1}$,
resulting in the personalized item $\bm{i}_n=\bm{i}_{n,0}$ that seamlessly complements $\bm{O}_n'$.

\vspace{3pt}
\noindent\textbf{$\bullet$ GOR.} 
Given designated categories $\{c_k\}_{k=1}^n$, category prompts $\{\bm{t}_{c_k}\}_{k=1}^n$, and user's interaction history $\{\bm{u}_{c_k}\}_{k=1}^n$, the target of GOR task is to synthesize a set of fashion products $\bm{O}=\{\bm{i}_k\}_{k=1}^n$ that exhibits internal compatibility and user-preferred aesthetics. 
Starting from a set of pure noises $\bm{O}_T=\{\bm{i}_{k,T}\}_{k=1}^n$ where $\bm{i}_{k,T}\sim\mathcal{N}(0,\bm{I})$, DiFashion computes the mutual condition $\{\bm{m}_{k,T}\}_{k=1}^n$ and history condition $\{\bm{h}_{c_k}\}_{k=1}^n$ via Eq.~(\ref{eq:mutual_cond}) and Eq.~(\ref{eq:hist_cond}), respectively. Next, it predicts the noise $\{\bm{\tilde{\epsilon}}_{\theta,\phi}(\bm{i}_{k,T},\bm{t}_{c_k},\bm{m}_{k,T},\bm{h}_{c_k},T)\}_{k=1}^n$ at step $T$ via Eq.~(\ref{eq:cfg_difashion}), and removes these noises $\bm{O}_T\rightarrow\bm{O}_{T-1}$. By repeating this process, 
DiFashion generates the final outfit $\bm{O}=\bm{O}_0$ that is visually appealing and satisfies the user's fashion taste.


\section{Experiments}
\label{sec:experiment}
We conduct extensive experiments to answer three questions:
\begin{itemize}[leftmargin=*]
    \item \textbf{RQ1:} How does DiFashion perform in PFITB and GOR tasks compared to both generative methods and retrieval-based OR methods, based on quantitative evaluation?
    \item \textbf{RQ2:} Can DiFashion surpass the baselines in human-involved qualitative evaluation?
    \item \textbf{RQ3:} How do the designs of DiFashion impact the performance, such as the mutual influence ratio $\eta$, the guidance scales $s_t$, $s_m$, and $s_h$, along with the MLP in the mutual encoder and the mutual and history conditions?
\end{itemize}

\begin{table}[t]
\setlength{\abovecaptionskip}{0cm}
\setlength{\belowcaptionskip}{0cm}
\caption{Statistics of two datasets.}
\label{table:dataset}
\begin{tabular}{lcccc}
\hline
                                      & \textbf{\#User}      & \textbf{\#Outfit}    & \textbf{\#Item}      & \textbf{\#Interaction}       \\ \hline
\multicolumn{1}{c}{\textbf{iFashion}} & 12,806                & 19,882                & 344,186               & 107,396               \\
\multicolumn{1}{c}{\textbf{Polyvore-U}} & 517                  & 33,906                & 119,202               & 33,908                \\ \hline
                                      & \multicolumn{1}{l}{} & \multicolumn{1}{l}{} & \multicolumn{1}{l}{} & \multicolumn{1}{l}{}
\end{tabular}
\vspace{-0.7cm}
\end{table}

\begin{table*}[t]
\setlength{\abovecaptionskip}{0.05cm}
\setlength{\belowcaptionskip}{0.1cm}
\caption{Quantitative performance comparison between DiFashion and generative baselines in both PFITB and GOR tasks. Baselines labeled with ``$\ast$" indicate pre-trained models, and ``Comp.'' and ``Per.'' denote compatibility and personalization, respectively. The best results are highlighted in bold, while the second-best results are underlined.}
\label{table:overall_gen}
\resizebox{\textwidth}{!}{
\begin{tabular}{l|cccccccc|ccccccc}
\hline
\textbf{\#iFashion} & \multicolumn{8}{c|}{\textbf{PFITB}}                                                                                        & \multicolumn{7}{c}{\textbf{GOR}}                                                                                      \\
\textbf{Methods}    & \textbf{FID$\downarrow$}& \textbf{IS$\uparrow$}    & \textbf{IS-acc$\uparrow$} & \textbf{CS$\uparrow$}    & \textbf{CIS$\uparrow$}   & \textbf{LPIPS$\downarrow$} & \textbf{Comp.$\uparrow$} & \textbf{Per.$\uparrow$}  & \textbf{FID$\downarrow$}   & \textbf{IS$\uparrow$}    & \textbf{IS-acc$\uparrow$} & \textbf{CS$\uparrow$}    & \textbf{CIS$\uparrow$}   & \textbf{LPIPS$\downarrow$} & \textbf{Per.$\uparrow$}  \\ \hline
\textbf{OutfitGAN}  & 202.60         & 9.52           & 0.19            & 18.07          & 16.88          & 0.63           & 0.03           & 21.48          & -              & -              & -               & -              & -              & -              & -              \\
\textbf{SD-v1.5$^\ast$}& 52.62          & 22.54          & 0.76            & 28.30          & 38.35          & 0.68           & 0.08           & 46.31          & 38.14          & 23.20          & 0.77            & 28.46          & 55.84          & 0.69           & 46.45          \\
\textbf{SD-v2$^\ast$}     & 54.13          & 21.66          & 0.71            & \textbf{29.78} & 38.66          & 0.70           & 0.04           & 46.52          & 40.14          & 22.19          & 0.74            & \textbf{29.94} & 56.00          & 0.71           & 46.60          \\
\textbf{SD-v1.5}    & {\ul 42.47}    & 26.76          & {\ul 0.83}      & 25.67          & 44.80          & {\ul 0.62}     & {\ul 0.46}     & {\ul 53.16}    & {\ul 28.31}    & 26.90          & {\ul 0.84}      & 25.62          & 63.59          & {\ul 0.63}     & {\ul 53.24}    \\
\textbf{SD-v2}      & 44.87          & 25.85          & 0.80            & 25.88          & 44.81          & 0.66           & 0.39           & 52.99          & 30.37          & 25.82          & 0.82            & 25.87          & 64.51          & 0.66           & 53.06          \\
\textbf{SD-naive}   & 44.38          & 25.45          & 0.80            & 26.15          & {\ul 44.93}    & 0.66           & 0.36           & 52.95          & 30.53          & 25.43          & 0.81            & 26.19          & {\ul 64.71}    & 0.67           & 52.95          \\
\textbf{ControlNet} & 42.74          & {\ul 27.76}    & 0.81            & {\ul 29.02}    & 41.93          & 0.67           & 0.16           & 49.90          & 29.41          & {\ul 28.49}    & 0.82            & {\ul 29.03}    & 60.96          & 0.68           & 49.91          \\
\rowcolor[HTML]{EAEAEA} 
\textbf{DiFashion}  & \textbf{34.06} & \textbf{29.99} & \textbf{0.90}   & 26.27          & \textbf{47.36} & \textbf{0.51}  & \textbf{0.58}  & \textbf{55.86} & \textbf{20.21} & \textbf{30.04} & \textbf{0.90}   & 26.10          & \textbf{67.35} & \textbf{0.54}  & \textbf{55.54} \\ \hline
\end{tabular}
}
\vspace{0.1cm}
\resizebox{\textwidth}{!}{
\begin{tabular}{l|cccccccc|ccccccc}
\hline
\textbf{\#Polyvore-U}   & \multicolumn{8}{c|}{\textbf{PFITB}}                                                                                        & \multicolumn{7}{c}{\textbf{GOR}}                                                                                            \\
\textbf{Methods}    & \textbf{FID$\downarrow$}   & \textbf{IS$\uparrow$}    & \textbf{IS-acc$\uparrow$} & \textbf{CS$\uparrow$}    & \textbf{CIS$\uparrow$}   & \textbf{LPIPS$\downarrow$} & \textbf{Comp.$\uparrow$} & \textbf{Per.$\uparrow$}  & \textbf{FID$\downarrow$}   & \textbf{IS$\uparrow$}          & \textbf{IS-acc$\uparrow$} & \textbf{CS$\uparrow$}    & \textbf{CIS$\uparrow$}   & \textbf{LPIPS$\downarrow$} & \textbf{Per.$\uparrow$}  \\ \hline
\textbf{OutfitGAN}  & 272.82         & 13.63          & 0.08            & 15.08          & 23.23          & 0.70           & 0.00           & 28.80          & -              & -                    & -               & -              & -              & -              & -              \\
\textbf{SD-v1.5$^\ast$}   & 51.34          & 17.10          & 0.73            & 27.46          & 40.56          & 0.67           & 0.70           & 51.05          & 42.59          & 16.95                & 0.73            & 27.51          & 51.68          & 0.68           & 50.99          \\
\textbf{SD-v2$^\ast$}     & 62.41          & 14.83          & 0.68            & \textbf{29.79} & 40.51          & 0.71           & 0.60           & 51.29          & 53.40          & 14.88                & 0.67            & \textbf{29.83} & 50.91          & 0.72           & 51.23          \\
\textbf{SD-v1.5}    & 53.72& 17.12          & 0.72& 24.31          & 45.71          & {\ul 0.64}     & {\ul 0.75}     & 58.20& 44.73& 17.24                & 0.72& 24.34          & 58.68          & {\ul 0.65}     & 58.16\\
\textbf{SD-v2}      & 58.13          & 15.59          & 0.67            & 24.49          & 46.15          & {\ul 0.64}     & 0.71           & 58.79          & 41.61          & 16.33                & 0.70            & 23.71          & 58.86          & 0.66           & 58.91          \\
\textbf{SD-naive}   & 58.22          & 15.45          & 0.66            & 24.23          & {\ul 47.01}    & 0.65           & 0.73           & {\ul 59.24}    & 49.04          & 15.48                & 0.67            & 24.31          & {\ul 59.18}    & 0.66           & {\ul 59.12}    \\
\textbf{ControlNet} & {\ul 44.31}    & {\ul 18.93}    & {\ul 0.77}      & {\ul 28.67}    & 43.79          & 0.68           & 0.73           & 55.44          & {\ul 36.05}    & \textbf{19.21}& {\ul 0.77}      & {\ul 28.72}    & 56.04          & 0.69           & 55.40          \\
\rowcolor[HTML]{EAEAEA} 
\textbf{DiFashion}  & \textbf{32.58} & \textbf{19.67} & \textbf{0.84}   & 25.53          & \textbf{48.82} & \textbf{0.56}  & \textbf{0.80}  & \textbf{61.44} & \textbf{23.47} & {\ul 18.95}& \textbf{0.83}   & 25.35          & \textbf{60.72} & \textbf{0.59}  & \textbf{61.16} \\ \hline
\end{tabular}
}
\vspace{-0.4cm}
\end{table*}
\subsection{Experimental Settings}
\subsubsection{\textbf{Datasets.}} We use two fashion datasets: 1) \textbf{iFashion}\footnote{https://github.com/wenyuer/POG.} includes pre-defined outfits and individual fashion products, along with users' click behaviors on these outfits and products; and 2) \textbf{Polyvore-U}\footnote{https://github.com/lzcn/Fashion-Hash-Net.} is collected from the Polyvore website with abundant pre-defined outfits and user-outfit interactions.

For the iFashion dataset, we only retain outfits with a length of four to save computing resources, focusing on basic outfit categories. Notably, DiFashion can effortlessly handle outfits with longer lengths and more diverse categories. 
For the Polyvore-U dataset, we choose Polyvore-519 due to its large user base and variant lengths of outfits. 
Here, items in outfits with a length of three are treated as interacted items, while outfits with four are considered interacted outfits. Since the Polyvore-U dataset only offers rough categories (top, bottom, shoes), we finetune a classifier Inception-V3~\cite{szegedy2016inception} using the iFashion dataset to predict fine-grained categories for Polyvore-U items. We only keep active users with at least five interacted outfits for both datasets, following~\cite{outfitgan} to split interacted outfits into training, validation, and testing sets with a ratio of 8:1:1. Dataset statistics are summarized in Table~\ref{table:dataset}.

\subsubsection{\textbf{Baselines.}} 
We compare DiFashion with competitive generative models on two tasks: 
    \textbf{1) OutfitGAN}~\cite{outfitgan} utilizes Generative Adversarial Network to generate a single fashion image for the FITB task.
    \textbf{2) SD-v1.5}\footnote{https://huggingface.co/runwayml/stable-diffusion-v1-5.} is a version of SD with a frozen CLIP ViT-L/14 text encoder for text prompts, trained on LAION-5B and ``LAION-Aesthetics V2 5+" datasets.
    \textbf{3) SD-v2}\footnote{https://huggingface.co/stabilityai/stable-diffusion-2-base.} is another version of SD trained on the LAION-5B dataset with a frozen OpenCLIP-ViT/H text encoder.
    \textbf{4) SD-naive} is an intuitive adaptation of SD for the GOR task, integrating mutual and history conditions through concatenation without the MLP in the mutual encoder. \textbf{5) ControlNet}~\cite{zhang2023controlnet} enhances a pre-trained DM by introducing spatial conditioning controls. It locks the pre-trained U-Net parameters and copies the weights of the encoding layers for finetuning.

Additionally, we also compare DiFashion with some retrieval-based OR methods under the retrieval setting:
    \textbf{1) Random} is a naive strategy that randomly samples fashion items for the PFITB and GOR tasks.
    \textbf{2) HFN}~\cite{lu2019hfn} is a POR method that models outfit compatibility and user preferences using content features and user representation with binary codes.
    \textbf{3) HFGN}~\cite{li2020hfgn} is another POR method using graphs to model outfit and user-outfit interactions.
    \textbf{4) BGN}~\cite{bai2019bgn} utilizes LSTM for bundle list recommendation. Modeling outfits as ordered sequences, it serves as a typical method for personalized outfit composition.
    \textbf{5) BGN-Trans}, designed by us, replaces the LSTM in BGN with a Transformer for personalized outfit composition, similar to~\cite{chen2019pog} for the same task.

\vspace{-0.1cm}
\subsubsection{\textbf{Evaluation Metrics.}} 
We compare DiFashion with the above baselines through the following quantitative metrics:
\vspace{-0.15cm}
\begin{itemize}[leftmargin=*]
    \item \textbf{Generative metrics.} We assess the fidelity of generated images using three widely used metrics: FID, IS, and CLIP Score (CS).
    For accurate measurement of IS, we finetune Inception-V3~\cite{szegedy2016inception} on the iFashion dataset. As DiFashion generates outfits with designated categories instead of random generation, we modify IS by replacing the edge distribution of generated image categories with a uniform distribution. The classification accuracy of generated images is also reported as IS-acc.
    \item \textbf{Fashion metrics.} We introduce three types of fashion metrics tailored for the PFITB and GOR tasks: 1) \textit{Similarity}: we calculate CLIP Image Score (CIS) via the cosine similarity of CLIP features between the generated images and ground-truth images for semantic similarity evaluation and adopt LPIPS~\cite{zhang2018lpips} for perceptual similarity evaluation. 2) \textit{Compatibility}: the compatibility evaluator in OutfitGAN~\cite{outfitgan} is employed to assess outfit compatibility. 3) \textit{Personalization}: we utilize CLIP to extract image features of user-interacted items, average them as user preference representation, and calculate the cosine similarity between user preference and generated images.
    \item \textbf{Retrieval accuracy.} For the PFITB task, we randomly sample four items from the item set to serve as negative samples, which are then combined with the ground-truth item to form the entire candidate set for each outfit. To calculate the retrieval accuracy, we ground each generated image to an existing item in the candidate set through the cosine similarity calculated by CLIP features.
    For the GOR task, the candidate sets are composed of all the items within each designated category.
\end{itemize}

\vspace{-0.2cm}
\subsubsection{\textbf{Implementation Details.}} We optimize generative models by selecting optimal checkpoints and hyper-parameters based on the average improvement ratio across four key metrics (FID, CIS, Compatibility, and Personalization) relative to the pre-trained SD-v2. SD-naive, ControlNet and our DiFashion are implemented on the latest Stable Diffusion SD-v2 for fair comparison. All the DM-based baselines and DiFashion have a fixed learning rate at $1e^{-5}$ and are finetuned on the iFashion dataset. For the Polyvore-U dataset, we finetune the model on top of the model trained on iFashion,
as the fine-grained categories of Polyvore-U are predicted from a classifier trained on iFashion. Other baselines are tuned according to their default settings.

In DiFashion, the hidden dimension of the mutual encoder is fixed at 256. We explore the mutual influence ratio $\eta$ in $\{0.05,0.10,0.15\}$ and search for the condition guidance scales $s_t$, $s_m$ and $s_h$ in $\{10.0,\cdots,15.0\}$, $\{2.0,\cdots,7.0\}$, and $\{2.0,\cdots,7.0\}$, respectively. During training, individual conditions are randomly masked with a ratio of $0.2$, and both the mutual and history conditions are simultaneously masked with a ratio of $0.3$. 
Training for DiFashion typically requires one A100 GPU day.

\begin{table*}[t]
\setlength{\abovecaptionskip}{0cm}
\setlength{\belowcaptionskip}{-0.1cm}
\caption{Comparison between DiFashion and retrieval-based models over the PFITB task. ``Comp.'' and ``Per.'' denote compatibility and personalization, respectively. The best results are highlighted in bold, while the second-best results are underlined.}
\label{table:fitb_retrieval}
\setlength{\tabcolsep}{3.8mm}{
\resizebox{\textwidth}{!}{
\begin{tabular}{l|cccccc|cccccc}
\hline
   & \multicolumn{6}{c|}{\textbf{iFashion}}                                                                                           & \multicolumn{6}{c}{\textbf{Polyvore-U}}                                                                                           \\
\textbf{Methods}             & \textbf{CS$\uparrow$}    & \textbf{CIS$\uparrow$}   & \textbf{LPIPS$\downarrow$} & \textbf{Comp.$\uparrow$} & \textbf{Per.$\uparrow$}  & \multicolumn{1}{l|}{\textbf{Retrieval$\uparrow$}} & \textbf{CS$\uparrow$}    & \textbf{CIS$\uparrow$}   & \textbf{LPIPS$\downarrow$} & \textbf{Comp.$\uparrow$} & \textbf{Per.$\uparrow$}  & \multicolumn{1}{l}{\textbf{Retrieval$\uparrow$}} \\ \hline
\textbf{Random}              & 13.79          & 48.89          & 0.42           & 0.25           & 50.50          & 0.17                                        & 13.90          & 54.18          & 0.40           & 0.45           & 57.88          & 0.22                                       \\
\textbf{HFN}                 & 18.44          & 69.98          & 0.24           & 0.55           & 56.79          & 0.51                                        & 15.59          & 64.36          & 0.31           & 0.55           & 60.31          & 0.39                                       \\
\textbf{HFGN}                & {\ul 21.64}    & {\ul 85.85}    & {\ul 0.12}     & {\ul 0.75}     & {\ul 62.25}    & {\ul 0.75}                                  & {\ul 20.07}    & \textbf{85.26} & \textbf{0.13}  & \textbf{0.90}  & {\ul 67.12}    & \textbf{0.74}                              \\
\rowcolor[HTML]{EAEAEA} 
\textbf{DiFashion} & \textbf{23.21} & \textbf{86.79} & \textbf{0.11}  & \textbf{0.78}  & \textbf{64.93} & \textbf{0.76}                               & \textbf{21.14} & {\ul 81.92}    & {\ul 0.17}     & {\ul 0.88}     & \textbf{68.49} & {\ul 0.66}                                 \\ \hline
\end{tabular}}}
\vspace{-0.2cm}
\end{table*}

\begin{table*}[ht]
\setlength{\abovecaptionskip}{0cm}
\setlength{\belowcaptionskip}{-0.1cm}
\caption{Comparison between DiFashion and retrieval-based models over the GOR task. ``Comp.'' and ``Per.'' denote compatibility and personalization, respectively. The best results are highlighted in bold, while the second-best results are underlined.}
\label{table:gor_retrieval}
\setlength{\tabcolsep}{5mm}{
\resizebox{0.9\textwidth}{!}{
\begin{tabular}{l|ccccc|ccccc}
\hline
\textbf{}          & \multicolumn{5}{c|}{\textbf{iFashion}}                                             & \multicolumn{5}{c}{\textbf{Polyvore-U}}                                              \\
\textbf{Methods}   & \textbf{CS$\uparrow$}    & \textbf{CIS$\uparrow$}   & \textbf{LPIPS$\downarrow$} & \textbf{Comp.$\uparrow$} & \textbf{Per.$\uparrow$}  & \textbf{CS$\uparrow$}    & \textbf{CIS$\uparrow$}   & \textbf{LPIPS$\downarrow$} & \textbf{Comp.$\uparrow$} & \textbf{Per.$\uparrow$}  \\ \hline
\textbf{Random}    & 23.79          & 69.13          & \textbf{0.45}  & 0.34           & 60.89          & 21.47          & 62.77          & {\ul 0.49}     & 0.70           & 65.82          \\
\textbf{BGN}       & 23.35          & 68.53          & {\ul 0.46}     & {\ul 0.62}     & 60.48          & 22.33          & \textbf{66.47} & \textbf{0.48}  & \textbf{0.95}  & 68.69          \\
\textbf{BGN-Trans} & {\ul 23.93}    & {\ul 69.87}    & \textbf{0.45}  & \textbf{0.75}  & {\ul 61.92}    & {\ul 22.60}    & {\ul 65.33}    & \textbf{0.48}  & {\ul 0.92}     & {\ul 68.75}    \\
\rowcolor[HTML]{EAEAEA} 
\textbf{DiFashion} & \textbf{25.51} & \textbf{70.95} & 0.47           & 0.52           & \textbf{63.74} & \textbf{24.36} & 65.01          & 0.50           & 0.80           & \textbf{69.94} \\ \hline
\end{tabular}
}}
\vspace{-0.35cm}
\end{table*}

\subsection{Quantitative Evaluation (RQ1)}

\subsubsection{\textbf{Comparison with Generative Models.}} We initially compare DiFashion with generative baselines in both PFITB and GOR tasks in Table~\ref{table:overall_gen}, revealing the following insights:
\begin{itemize}[leftmargin=*]
    \item DM-based models outperform OutfitGAN in image fidelity for the PFITB task, showcasing the exceptional capabilities of DMs in fashion image synthesis. Despite integrating a compatibility module, OutfitGAN generates lower-quality images, resulting in inferior compatibility. Besides, OutfitGAN struggles with the GOR task due to inherent limitations in its model structure.
    \item After finetuning on each dataset, both SD-v1.5 and SD-v2 exhibit enhanced generation fidelity, resulting in improved similarity (CIS, LPIPS), compatibility, and personalization. The decline in CS might stem from the category prompts lacking adequate fashion details, leading to reduced alignment between generated images and prompts. Besides, SD-v1.5 surpasses SD-v2 in certain metrics, possibly due to the differences in pre-training datasets.
    \item SD-naive and ControlNet integrate mutual and history conditions, while their performance remains unsatisfactory. SD-naive directly incorporates these conditions without additional model designs to capture useful information, potentially introducing extra noise into the U-Net. As for ControlNet, it freezes the pre-trained U-Net and only finetunes two copies of U-Net encoding layers to handle conditions. However, the fixed U-Net might limit the generalization ability of DMs to model new data distributions while integrating multiple conditions.
    \item DiFashion exhibits superior performance across most evaluation metrics on two datasets, confirming its excellence in both PFITB and GOR tasks. The improvements in compatibility and personalization highlight its effectiveness in modeling compatibility information and user preferences.
    \item Performance in PFITB and GOR tasks demonstrates a similar trend, as PFITB could be viewed as a simplified version of GOR. The compatibility results in the GOR task are close to zero for all methods and thus omitted to save space. These poor results might stem from the potential inconsistencies between generated and real-world images, impacting the effectiveness of the compatibility evaluator trained on real-world images. Generating all outfit images in the GOR task introduces additional complexity, making quantitative compatibility evaluation more challenging than in the PFITB task.
    Human-involved qualitative evaluation on compatibility is conducted in Section~\ref{sec:human_evaluation}.
\end{itemize}

\begin{figure*}[t]
\setlength{\abovecaptionskip}{-0cm}
\setlength{\belowcaptionskip}{-0.45cm}
\centering
\includegraphics[height=1.55in]{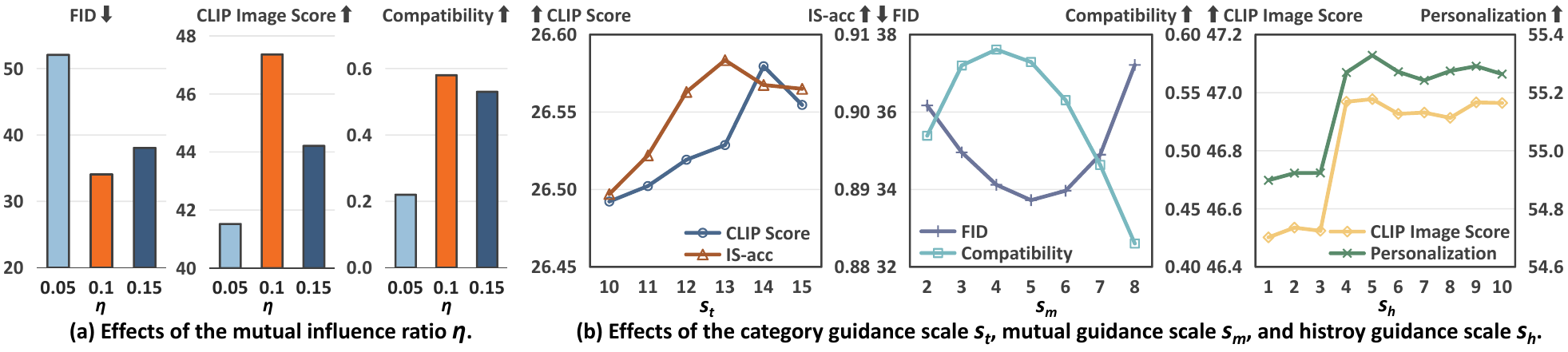}
\caption{Effects of the mutual influence ratio and three guidance scales.}
\label{fig:hyperparam}
\end{figure*}

\vspace{-0.1cm}
\subsubsection{\textbf{Comparison with Retrieval-based Models.}}
We retrieve the proximal existing fashion products for the images generated by DiFashion and conduct a comparative analysis with retrieval-based baselines. Notably, DiFashion is evaluated in a zero-shot scenario, as it is trained without exposure to negative outfits.
\begin{itemize}[leftmargin=*]
    \item Table~\ref{table:fitb_retrieval} presents the performance comparison in the PFITB task, showcasing that DiFashion achieves comparable performance to retrieval-based models across diverse metrics. Particularly, DiFashion surpasses all the baselines on the iFashion dataset, emphasizing its effectiveness in capturing compatibility information and user preferences. This success affirms the practical feasibility of implementing DiFashion-generated images in real-world scenarios through retrieval mechanisms.
    \item We present retrieval results for the GOR task in Table~\ref{table:gor_retrieval}, where DiFashion still achieves comparable performance with baselines. Retrieval accuracy is omitted here due to the difficulty of accurately aligning generated outfits with the ground truth, given the diverse combination of existing items for outfit composition. Besides, DiFashion-retrieved outfits are less compatible than the PFITB task, possibly due to visual distinctions between generated fashion products and existing ones, posing challenges in achieving a perfect match. The GOR task involves retrieving four items, in contrast to the single-item retrieval in PFITB, potentially impacting compatibility significantly.

\end{itemize}

\begin{table}[t]
\setlength{\abovecaptionskip}{0.05cm}
\setlength{\belowcaptionskip}{0cm}
\caption{The human-involved qualitative evaluation results, where ``$\pm$'' denotes 95\% confidence interval. DiFashion is consistently preferred ($\ge50\%$) over the baselines across all evaluation metrics for both PFITB and GOR tasks.}
\label{table:human_eval}
\resizebox{0.48\textwidth}{!}{
\begin{tabular}{cc|ccc}
\hline
\multicolumn{2}{c|}{\textbf{DiFashion}}                                  & \textbf{Fidelity} & \textbf{Compatibility} & \textbf{Personalization} \\ \hline \xrowht{10pt}
\multirow{2}{*}{\textbf{PFITB}} & \multicolumn{1}{c|}{\textbf{SD-v1.5}} & 64.08$^{\pm3.08\%}$     & 60.44$^{\pm2.42\%}$           & 68.32$^{\pm3.47\%}$             \\
                                & \multicolumn{1}{c|}{\textbf{SD-v2}}   & 70.04$^{\pm4.16\%}$      & 57.48$^{\pm1.90\%}$           & 66.40$^{\pm3.39\%}$             \\ \hline \xrowht{10pt}
\multirow{2}{*}{\textbf{GOR}}   & \multicolumn{1}{c|}{\textbf{SD-v1.5}} & 61.56$^{\pm1.93\%}$      & 61.20$^{\pm2.00\%}$           & 60.80$^{\pm2.57\%}$              \\
                                & \multicolumn{1}{c|}{\textbf{SD-v2}}   & 66.52$^{\pm2.15\%}$      & 60.56$^{\pm1.88\%}$           & 63.72$^{\pm1.95\%}$             \\ \hline
\end{tabular}
}
\vspace{-0.45cm}
\end{table}

\vspace{-0.18cm}
\subsection{\large{Human-involved Qualitative Evaluation (RQ2)}}
\label{sec:human_evaluation}
To evaluate the qualitative performance of DiFashion in fidelity, compatibility, and personalization, we conduct a human evaluation on Amazon Mechanical Turk\footnote{https://www.mturk.com/.}, 
comparing it with two competitive baselines (SD-v1.5 and SD-v2) using binary-choice tests on the iFashion dataset.
The evaluation covers both PFITB and GOR tasks, each with 50 cases. For fidelity assessment, we present generated images/outfits with the question: ``Which image/outfit provides a more realistic and complete portrayal of fashion products with a clean background?''. To evaluate compatibility, we display entire outfits with questions: ``Which fashion image is more compatible with this incomplete outfit?'' for PFITB, and ``Which outfit is more compatible?'' for the GOR task. In the personalization evaluation, we show up to 5 user-interacted items per designated category for PFITB, and up to 2 for the GOR task. The question posed is, ``Considering someone's past liked fashion products, please select the option that comes closest to his/her preference.'' 
We collect 50 responses for each questionnaire. 
Table~\ref{table:human_eval} reports the selection proportions of DiFashion in each binary-choice test, along with 95\% confidence intervals. DiFashion consistently outperforms ($\ge50\%$) the baselines across all qualitative evaluation metrics, highlighting its superiority in meeting the three criteria of the GOR task.

\begin{table}[t]
\setlength{\abovecaptionskip}{-0cm}
\setlength{\belowcaptionskip}{0.1cm}
\caption{Performance of DiFashion with (w/) and without (w/o) the MLP in the mutual encoder.}
\label{table:mlp_ablation}
\begin{tabular}{lcccc}
\hline
\textbf{Variants} & \textbf{FID$\downarrow$} & \textbf{CIS$\uparrow$} & \textbf{Comp.$\uparrow$} & \textbf{Per.$\uparrow$} \\ \hline
\textbf{w/o MLP}  & 100.56       & 31.72        & 0.16           & 38.99         \\
\textbf{w/ ~~MLP}   & 34.06        & 47.36        & 0.58           & 55.86         \\ \hline
\end{tabular}
\vspace{-0.45cm}
\end{table}

\vspace{-0.2cm}
\subsection{In-depth Analysis (RQ3)}
In this section, we focus on the impact of various designs in DiFashion and only report the results of the PFITB task on the iFashion dataset, omitting the results of the GOR task or on the Polyvore-U dataset to save space.

\vspace{-0.1cm}
\subsubsection{\textbf{Hyper-parameter Analysis.}}
\leavevmode

\vspace{3pt}
\noindent\textbf{$\bullet$ Effects of the mutual influence ratio $\eta$.} 
The mutual conditions, controlled by $\eta$, guide the generation process in DiFashion. 
We vary $\eta\in\{0.05,0.10,0.15\}$ during training and report the results over three highly related metrics
in Figure~\ref{fig:hyperparam}(a), which reveal the following insights: 1) A too-small $\eta$ weakens compatibility 
guidance, causing declines in fidelity, similarity to ground truth, and compatibility. 2) Conversely, with a too-large $\eta$, the mutual condition dominates the generation process, diminishing the influence of other conditions and resulting in inferior performance.

\vspace{3pt}
\noindent\textbf{$\bullet$ Effects of guidance scales $s_t$, $s_m$ and $s_h$.} We vary three guidance scales during inference and present the results in Figure~\ref{fig:hyperparam}(b), focusing on the two most relevant metrics respectively. Observations from the figure reveal that: 1) Increasing the category guidance scale $s_t$ boosts both CLIP Score and IS-acc, indicating improved consistency of generated images with designated categories. 2) As the mutual guidance scales $s_m$ increases, performance in FID and Compatibility initially rises, validating the effectiveness of the mutual condition in compatibility modeling. However, excessive enlargement weakens influences from other conditions, degrading fidelity and compatibility. Thus, careful selection of an appropriate $s_m$ for the mutual condition is crucial. 3) Increasing the history guidance scale $s_h$ leads to higher scores in personalization and similarity to ground truth, indicating enhanced consistency of generated images with user preferences.

\begin{figure}[t]
\setlength{\abovecaptionskip}{0.cm}
\setlength{\belowcaptionskip}{-0.6cm}
\centering
\includegraphics[height=1.5in]{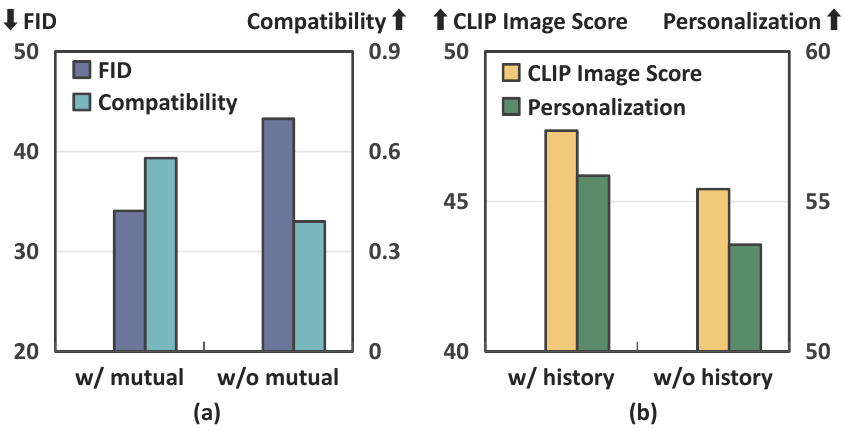}
\caption{Effects of the mutual and history conditions. ``w/'' and ``w/o'' denote ``with'' and ``without'', respectively.}
\label{fig:ablation_condition}
\end{figure}

\vspace{-0.1cm}
\subsubsection{\textbf{Ablation Study.}}
\leavevmode

\vspace{3pt}
\noindent\textbf{$\bullet$ Effects of the MLP in the mutual encoder.} 
To assess the impact of the MLP in the mutual encoder, we compare the performance of DiFashion with and without the MLP, as shown in Table~\ref{table:mlp_ablation}, focusing on four key metrics.
The declined performance validates the effectiveness of the MLP in capturing compatibility information for guiding the generation process. The absence of the MLP introduces excessive noise in the mutual condition, hindering compatibility guidance and even influencing guidance from other conditions, leading to inferior performance across all metrics.

\vspace{3pt}
\noindent\textbf{$\bullet$ Effects of the mutual and history conditions.} To explore the impact of history and mutual conditions, we perform additional experiments excluding mutual or history conditions during training. The results in Figure~\ref{fig:ablation_condition} reveal the impact on relevant metrics. In subfigure (a), the lack of the mutual condition results in a decrease in both fidelity and compatibility. This underscores the efficacy of the mutual condition in capturing compatibility information, which not only enhances compatibility but also enriches the generation process with additional details, resulting in high-fidelity image generation. In subfigure (b), the absence of the history condition leads to a decline in both personalization and similarity to ground truth, affirming the effectiveness of the history condition in capturing users' fashion tastes.

\vspace{-0.2cm}
\subsection{Case Study}
In this section, we present some examples of fashion products and outfits created by DiFashion and further explore the capabilities of DiFashion in generalized GOR task.

\begin{figure}[t]
\setlength{\abovecaptionskip}{0.1cm}
\setlength{\belowcaptionskip}{-0.4cm}
\centering
\includegraphics[scale=0.34]{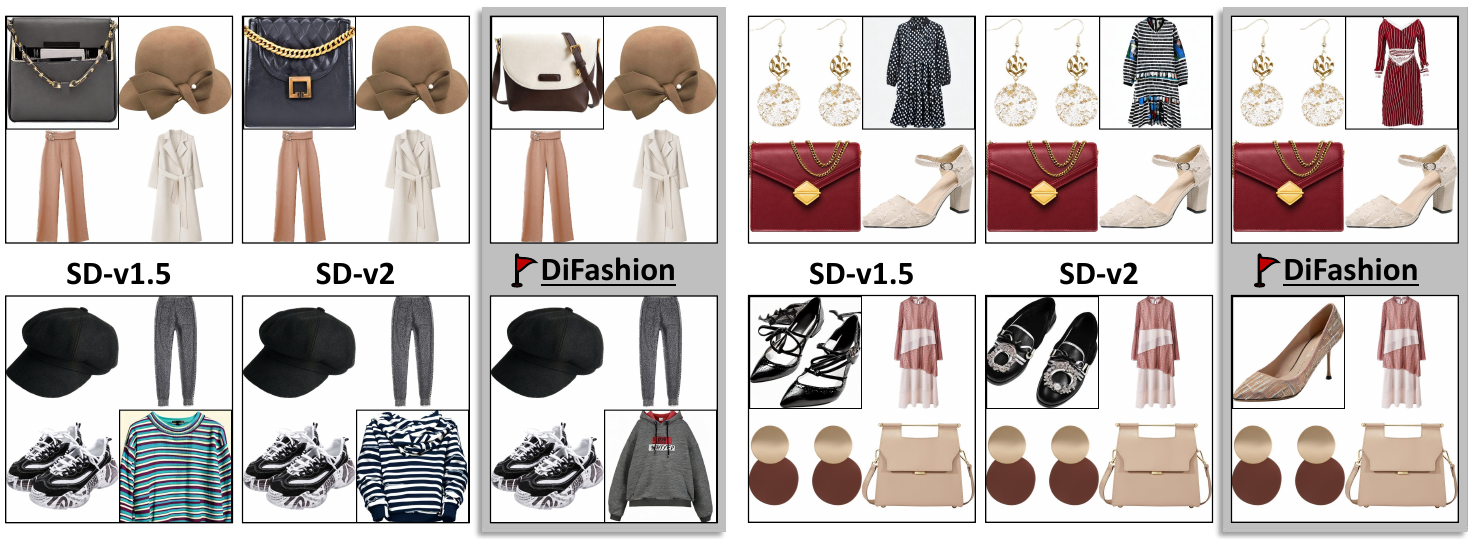}
\caption{Examples of generated images (enclosed in small black squares within each outfit) for the PFITB task.}
\label{fig:fitb_case}
\end{figure}

\vspace{3pt}
\noindent\textbf{$\bullet$ DiFashion for the PFITB task.} In Figure~\ref{fig:fitb_case}, we showcase the generated fashion products for the PFITB task, each enclosed in a small black square within the outfit. The comparison between DiFashion and the two baselines highlights DiFashion's capability to generate fashion images with superior fidelity and compatibility.

\begin{figure}[t]
\setlength{\abovecaptionskip}{0cm}
\setlength{\belowcaptionskip}{-0.5cm}
\centering
\includegraphics[scale=0.42]{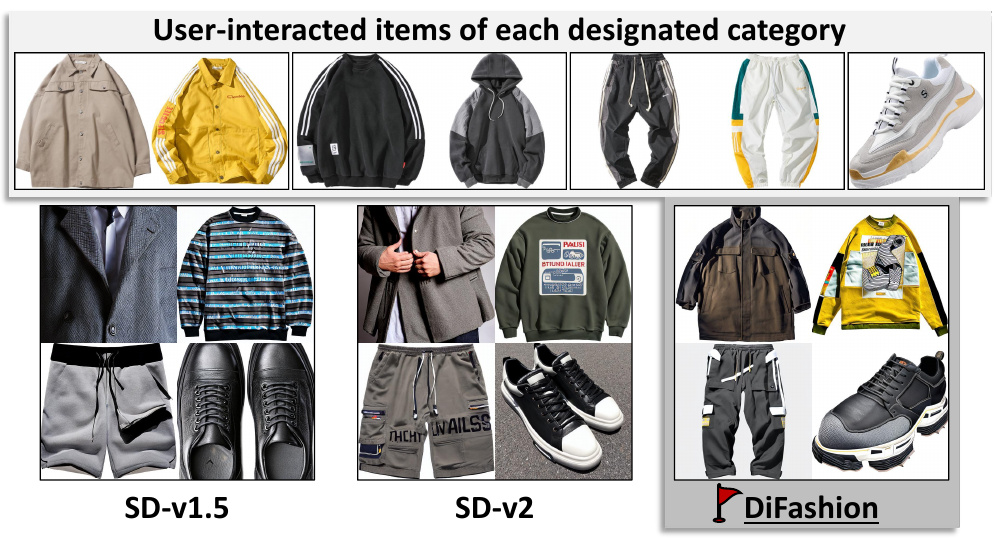}
\caption{An example of generated outfits for the GOR task, along with user-interacted items of each category.}
\label{fig:gor_case}
\end{figure}

\vspace{3pt}
\noindent\textbf{$\bullet$ DiFashion for the GOR task.} Figure~\ref{fig:gor_case} presents example outfits alongside the user's interaction history for the GOR task, comparing DiFashion with SD-v1.5 and SD-v2. The comparison reveals DiFashion's superiority in meeting the three criteria of the GOR task: 1) \textit{High fidelity:} DiFashion excels in generating more realistic fashion images with clean backgrounds; 2) \textit{Comaptibility:} DiFashion-generated outfit is more harmonious in occasion, season, and style; 3) \textit{Personalization:} DiFashion incorporates user's interaction history to generate a more personalized outfit aligning with user preferences in color, pants-length, and sports-style.

\begin{figure}[t]
\setlength{\abovecaptionskip}{0cm}
\setlength{\belowcaptionskip}{-0.35cm}
\centering
\includegraphics[scale=0.255]{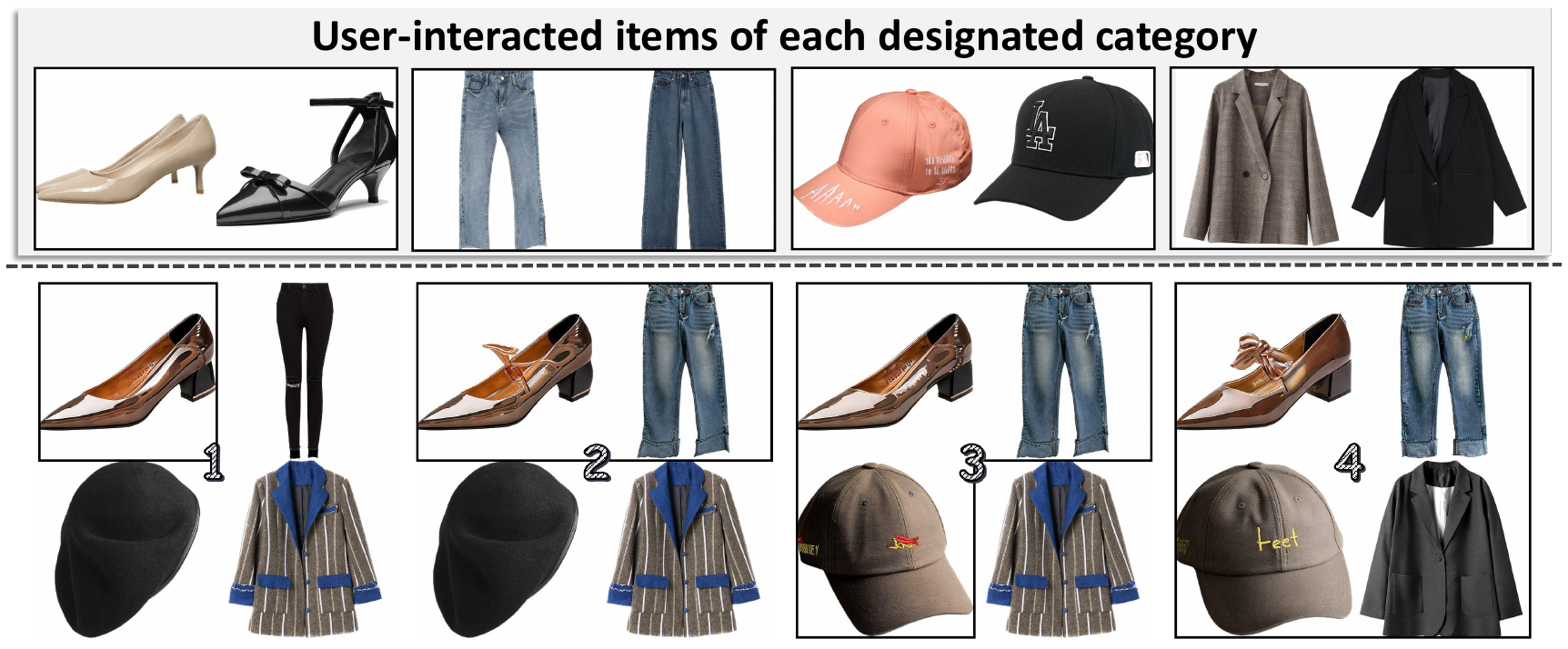}
\caption{DiFashion for generalized GOR task. Given user-interacted items and an incomplete outfit (including an empty set), DiFashion can generate an arbitrary quantity of personalized fashion items to compose an appealing outfit.}
\label{fig:generalized_gor}
\end{figure}

\vspace{3pt}
\noindent\textbf{$\bullet$ DiFashion for generalized GOR task.} Considering the PFITB task as a simplified GOR task, we can expand the GOR task into a more inclusive form: when presented with user interaction history and an incomplete outfit (including an empty set), the generalized GOR task aims to synthesize an arbitrary quantity of personalized fashion products to compose a visually compatible outfit. As shown in Figure~\ref{fig:generalized_gor}, DiFashion is capable of the generalized GOR task, creating compatible outfits that align with users' preferences.

\vspace{-0.1cm}

\section{Related Work}
\label{sec:related_work}

\noindent\textbf{$\bullet$ Outfit recommendation.} To recommend compatible outfits to align with users' fashion tastes, outfit recommendation emphasizes both compatibility and user preferences modeling~\cite{ding2023fashionsurvey,sarkar2023outfittransformer}. Early research views outfits as the minimum unit and conducts outfit-level retrieval for recommendation~\cite{li2020hfgn,lu2019hfn,hu2015cfrec,lin2020outfitnet,dong2019capsule}, limited in quantity and diversity of pre-defined outfits. Personalized outfit composition has subsequently emerged, employing item-level retrieval to compose outfits for specific users~\cite{han2017bilstm,chen2019pog,ding2023og,li2019coherent}. 
Whether retrieving at the outfit or item level, current methods are constrained by existing fashion products, struggling to satisfy users' diverse fashion needs~\cite{yu2019personalized,jain2019text}. 
In contrast, GOR generates new personalized products for outfit composition that could be retrieved or customized for practical implementation.

\vspace{3pt}
\noindent\textbf{$\bullet$ Generative recommendation.} Traditional recommender systems~\cite{liang2018variational,he2020lightgcn,rendle2009bpr} typically rely on retrieving items from existing datasets for personalized recommendation. The rise of powerful generative models~\cite{dhariwal2021diffusion,rombach2022high,brown2020language,ouyang2022training} enables recommender systems to integrate AIGC, moving forward to a more personalized paradigm to satisfy users' diverse content needs~\cite{wang2023generec}. However, existing works~\cite{wang2023diffrec,yang2023dreamrec,Alshehri2022news,liu2023once} mainly utilize generative models to enhance traditional recommendation. In contrast, GeneRec~\cite{wang2023generec} introduces the generative recommendation paradigm that incorporates generative models, such as DMs and Large Language Models (LLMs) to modify existing content or directly generate new content. The generated content is then integrated into the existing item corpus for recommendation. In this paradigm, CG4CTR~\cite{yang2024genctr} utilizes DMs for personalized advertisement generation based on user attributes, modifying the background of existing product images to meet user preferences. Different from CG4CTR, we propose an innovative generative task
in the Fashion domain -- Generative Outfit Recommendation, along with a novel model named DiFashion, aiming to generate new outfits to satisfy users' diverse fashion tastes.

\vspace{3pt}
\noindent\textbf{$\bullet$ Diffusion models for fashion.} Prior studies have explored the application of DMs in the fashion domain~\cite{cao2023image,kong2023leveraging}, with a focus on designer-centric fashion image synthesis. 
For example, SGDiff~\cite{sun2023sgdiff} and FashionDiff~\cite{yan2023fashiondiff} employ diverse elements (\eg sketch, color, texture) to generate fashion designs,
while MGD~\cite{baldrati2023multimodal} is proposed for multimodal-conditioned fashion image editing. Despite these advancements, utilizing DMs for user-centric fashion image synthesis remains underexplored. Addressing this gap, DiFashion incorporates user interaction history to generate multiple tailored fashion images with internal compatibility for outfit composition, contributing to a highly personalized fashion landscape.

\section{Conclusion and Future Work}
\label{sec:conclusion}

In this work, we introduced Generative Outfit Recommendation, a novel task dedicated to synthesizing outfits with three key criteria: high fidelity, compatibility, and personalization. To achieve this, we proposed DiFashion, a generative outfit recommender model that employs three specific conditions to guide the parallel generation of multiple fashion images. Empirical results on two datasets validated the superiority of DiFashion in PFITB and GOR tasks.

This work introduces a new direction for outfit recommendation, leading to many promising ideas that deserve further exploration: 1) it is interesting to incorporate more diverse categories and more detailed controls over fashion attributes for the GOR task; 2) designing better strategies to encode compatibility and personalized information and integrate these multimodal conditions into DMs is a meaningful topic; 3) recognizing the evolving fashion trends in the GOR task is important, especially in the context of fast fashion~\cite{ding2021leveraging,ma2020trend}; and 4) leveraging the powerful semantic comprehension ability of LLMs through multi-round interactions with users to derive explicit instructions for user-guided GOR tasks holds promise for future research~\cite{bao2023bi,lin2023multi}.


{
\tiny
\bibliographystyle{ACM-Reference-Format}
\balance
\bibliography{bibfile}
}

\end{document}